\documentclass[
  aip,
  pop,
  amsmath,amssymb,
  reprint,
]{revtex4-1}
\usepackage{graphicx,color}
\usepackage{amsmath}
\usepackage{natbib}
\usepackage{epsfig}
\begin{document}

\title{On the long-waves dispersion in Yukawa systems}

\author{Sergey A. Khrapak}
\affiliation{Aix-Marseille-Universit\'{e}, CNRS, Laboratoire PIIM, UMR 7345, 13397 Marseille cedex 20, France}
\affiliation{Forschungsgruppe Komplexe Plasmen, Deutsches Zentrum f\"{u}r Luft- und Raumfahrt,
Oberpfaffenhofen, Germany}
\affiliation{Joint Institute for High Temperatures, Russian Academy of Sciences, Moscow, Russia}

\author{Boris Klumov}
\affiliation{Aix-Marseille-Universit\'{e}, CNRS, Laboratoire PIIM, UMR 7345, 13397 Marseille cedex 20, France}
\affiliation{Joint Institute for High Temperatures, Russian Academy of Sciences, Moscow, Russia}
\affiliation{L.D. Landau Institute for Theoretical Physics, Russian Academy of Sciences, Moscow, Russia}

\author{L\'{e}na\"{i}c Cou\"{e}del}
\affiliation{Aix-Marseille-Universit\'{e}, CNRS, Laboratoire PIIM, UMR 7345, 13397 Marseille cedex 20, France}

\author{Hubertus M. Thomas}
\affiliation{Forschungsgruppe Komplexe Plasmen, Deutsches Zentrum f\"{u}r Luft- und Raumfahrt,
Oberpfaffenhofen, Germany}

\begin{abstract}
A useful simplification of the quasilocalizded charge approximations (QLCA) method to calculate the dispersion relations in strongly coupled Yukawa fluids is discussed.
In this simplified version, a simplest possible model radial distribution function, properly related to the thermodynamic properties of the system, is used. The approach demonstrates good agreement with the dispersion relations obtained using the molecular dynamics simulations and the original QLCA in the long-wavelength regime.   
\end{abstract}

\pacs{52.27.Lw, 52.27.Gr, 52.35.Fp}
\date{\today}
\maketitle

\section{Introduction}

The quasilocalized charge approximation (QLCA) was originally proposed by Kalman and Golden\cite{KaGo1990} as a powerful formalism for the analysis of the dielectric response tensor and collective mode dispersion in strongly coupled Coulomb liquids.
The approach is based on a microscopic model in which the charges are quasilocalized on a short-time scale in local
potential fluctuations, for a review see Ref.~\onlinecite{GoldenPoP2000}. In last decades the QLCA approach has been successively applied to various systems of strongly interacting particles to describe wave dispersion relations. In particular, this includes two-dimensional (2D) and three-dimensional (3D) one-component-plasma,~\cite{GoldenPoP2000}  2D and 3D Yukawa systems, mainly in the context of complex (dusty) plasmas,~\cite{RosenbergPRE1997,KalmanPRL2000,OhtaPRL2000,KalmanPRL2004,DonkoJPCM2008} classical 2D dipole systems,~\cite{GoldenPRB2008,GoldenPRE2010} and 3D dusty plasma with Lennard-Jones-like interactions.~\cite{RosenbergCPP2015}

Technically, for a given interaction potential, the QLCA approach requires as an input the equilibrium radial distribution function (RDF) $g(r)$, characterizing the spatial order in the system of particles. The latter can be obtained via various integral equation schemes or via the direct molecular dynamics (MD) or Monte Carlo (MC) simulations. Although, there are no principle difficulties in both these approaches, they remain relatively resource consuming. The purpose of this paper is to demonstrate that to describe the long-wavelength dispersion relations in strongly coupled Yukawa fluids, the accurate knowledge of $g(r)$ is unnecessary. The main effect of strong coupling can be accounted for by using a simple excluded volume consideration. A simplest possible toy $g(r)$ allows us to reproduce the long-wave dispersion curves with a reasonable good accuracy.     

In Yukawa systems, which are of some relevance in the context of colloidal suspensions and complex (dusty) plasmas,~\cite{IvlevBook,FortovUFN_2004,FortovRev}  the particles are interacting via the repulsive potential of the form $V(r)=(Q^2/r)\exp(-r/\lambda)$,
where $Q$ is the particle charge, $\lambda$ is the screening length, and $r$ is the distance between a pair of particles. These systems are conveniently characterized by two dimensionless parameters, which are  
the coupling parameter $\Gamma=Q^2/aT$, and the screening parameter $\kappa=a/\lambda$. Here $T$ is the system temperature (in energy units), $n$ is the particle density, and $a=(4\pi n/3)^{-1/3}$ is the characteristic inter-particle separation (Wigner-Seitz radius). The phase behavior of Yukawa systems is relatively well understood.~\cite{RobbinsJCP1988,HamaguchiPRE1997} For $\Gamma e^{-\kappa}\ll 1$ a weakly coupled gaseous regime is realized. As $\Gamma$ increases, the system shows a transition to the strongly coupled fluid regime. When $\Gamma$ increases further, the system crystallizes either into body-centered-cubic (bcc) or into the face-centered-cubic (fcc) lattice (bcc is thermodynamically favorable at weak screening, i.e. lower $\kappa$). The values of $\Gamma_{\rm m}(\kappa)$, corresponding to the fluid-crystal transition, have been tabulated;~\cite{HamaguchiPRE1997} relatively accurate fits are also available.~\cite{VaulinaJETP2000,VaulinaPRE2002,KhrapakPRL2009} For even higher $\Gamma$, the glass transition is predicted, with the glass-transition line almost parallel to the melting line in the extended region of the phase diagram.~\cite{YazdiPRE2014} In this study we focus on the strongly coupled fluid regime, characterized by $\Gamma\lesssim \Gamma_{\rm m}$.    

\section{Quasilocalized charge approximation and its simplified version}

The dispersion relations in the QLCA approach read
\begin{equation}
\begin{aligned}\label{disp}
\omega_{L}^2=\omega_0^2(q)+D_L(q),\\
\omega_T^2=D_T(q),
\end{aligned}
\end{equation}
where $q=ka$ is the reduced wave number and the subscripts ``$L$'' and ``$T$'' stand for longitudinal and transverse modes, respectively. The term $\omega_0(q)$ corresponds to the longitudinal dispersion relation of non-correlated particles (weak-coupling limit),
\begin{equation}\label{w_0}
\omega_0^2(q)=\frac{\omega_{\rm p}^2 q^2}{q^2+\kappa^2},
\end{equation}
where $\omega_{\rm p}=\sqrt{4\pi Q^2 n/m}$ is the plasma frequency associated with the charged particle component and $m$ is the particle mass. In the context of complex (dusty) plasmas, this mode is known as the dust-acoustic-wave (DAW).~\cite{Rao1990} The respective projections of the QLCA dynamical matrix $D_{L}(q)$ and $D_T(q)$ are the functions of the equilibrium $g(r)$:~\cite{GoldenPoP2000,KalmanPRL2000, DonkoJPCM2008}
\begin{equation}\label{D}
D_{L/T}=\omega_{\rm p}^2\int_0^{\infty}\frac{dr}{r}\left[g(r)-1\right]{\mathcal K}_{L/T}(qr,\kappa r),
\end{equation}
with
\begin{equation}\label{K_L}
\begin{aligned}
{\mathcal K}_{L}(x,y)=-e^{-y}\left[\left(2+2y+\frac{2}{3}y^2\right)\right. \\ \left. \left(\frac{\sin x}{x}+3\frac{\cos x}{x^2}-3\frac{\sin x}{x^3}\right)+\frac{1}{3}y^2\left(\frac{\sin x}{x}-1\right)\right]
\end{aligned}
\end{equation}
and
\begin{equation}
\begin{aligned}\label{K_T}
{\mathcal K}_{T}(x,y)=e^{-y}\left[\left(1+y+\frac{1}{3}y^2\right)\right. \\ \left. \left(\frac{\sin x}{x}+3\frac{\cos x}{x^2}-3\frac{\sin x}{x^3}\right)-\frac{1}{3}y^2\left(\frac{\sin x}{x}-1\right)\right].
\end{aligned}
\end{equation}
The dimensionless distance $r$ in Eq.~(\ref{D}) and throughout the paper is expressed in units of $a$. 
When $g(r)$ is known, Eqs.~(\ref{disp}) -- (\ref{K_T}) allow us to calculate the dispersion relations of Yukawa systems in the QLCA approach.

The equilibrium RDF, $g(r)$, is also related to important thermodynamic quantities of the system such as energy and pressure. For pairwise interactions they can be expressed in terms of the integrals over $g(r)$, which are known as the energy and pressure (or virial) equations.~\cite{HansenBook} Since the dispersion relations and thermodynamic properties depend only on the integral of $g(r)$ it is not very unreasonable to presume that if a simple model form is chosen, which describes reasonably well the thermodynamic properties of the system, it will also allow to estimate the dispersion relations of this system. We shall now demonstrate that this is indeed a reasonable assumption, provided the long-wavelengths (longer than the mean interparticle separation) are of main interest.

To further pursue the link between dispersion properties and thermodynamics we chose the most simple step-wise toy model for $g(r)$, that is $g(r)=1$ for $r>R$ and $g(r)=0$ otherwise. Here $R$ characterizes an excluded volume around each particle due to strong (repulsive) inter-particle interactions. For this model $g(r)$, the integration in Eq.~(\ref{D}) can be performed analytically. The resulting dispersion relation of the longitudinal mode is
\begin{equation}\label{L1}
\begin{aligned}
\omega_L^2=\omega_{\rm p}^2e^{-R\kappa}\left[\left(1+R\kappa\right)\left(\frac{1}{3}-\frac{2\cos Rq}{R^2q^2}+\frac{2\sin Rq}{R^3q^3} \right) \right. \\ \left. -\frac{\kappa^2}{\kappa^2+q^2}\left(\cos Rq+\frac{\kappa}{q}\sin Rq \right)\right].
\end{aligned}
\end{equation}  
Similar, for the transverse mode we get
\begin{equation}\label{T1}
\omega_T^2=\omega_{\rm p}^2e^{-R\kappa}\left(1+R\kappa\right)\left(\frac{1}{3}+\frac{\cos Rq}{R^2q^2}-\frac{\sin Rq}{R^3q^3} \right).
\end{equation}
The remaining step is to find an appropriate model for the dimensionless parameter $R(\kappa,\Gamma)$ and to verify whether the proposed simplification can deliver reasonable results. 

Substituting the same step-wise RDF into the energy equation we easily obtain for the excess energy per particle in units of the system temperature
\begin{equation}\label{u_ex}
u_{\rm ex}=\frac{2\pi n a^3}{T}\int_0^{\infty}r^2V(r)g(r)dr=\frac{3\Gamma}{2\kappa^2}\left(1+R\kappa\right)e^{-R\kappa}.
\end{equation}        
Similarly, for the reduced excess pressure we get
\begin{equation}
\begin{aligned}\label{p_ex}
p_{\rm ex}= -\frac{2\pi n a^3}{3T}\int_0^{\infty}r^3V'(r)g(r)dr=\\
\frac{\Gamma}{2\kappa^2}\left(3+3R\kappa+R^2\kappa^2\right)e^{-R\kappa}.
\end{aligned}
\end{equation}
The effective radius of the exclusion sphere, $R$, can then be obtained from the solution of either Eq.~(\ref{u_ex}) or (\ref{p_ex}). Naturally, since the toy model for $g(r)$ is used, the result for $R$ somewhat depends on whether the energy or pressure route is used. The corresponding quantities will be referred to as $R_u$ and $R_p$, respectively. 
      
\begin{figure}
\includegraphics[width=7.5cm]{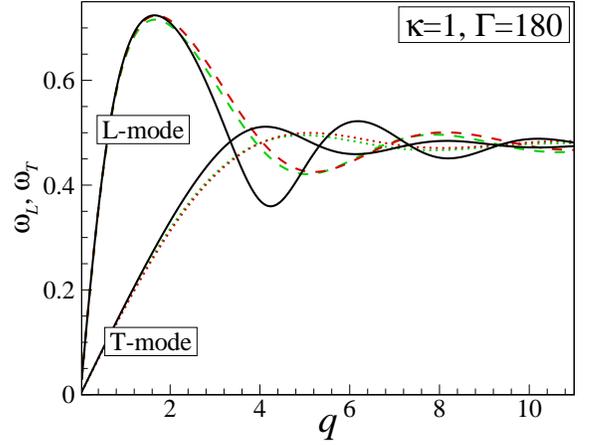}
\caption{(color online) The dispersion relations of the longitudinal (L-mode) and transverse (T-mode) waves in strongly coupled Yukawa fluid, characterized by $\kappa=1.0$ and $\Gamma=180$ (frequency is in units of the plasma frequency $\omega_{\rm p}$). The solid curves correspond to the conventional QLCA with the radial distribution function obtained via the direct MD simulations. The dashed (L-mode) and dotted (T-mode) curves correspond to the simplified version of QLCA of Eq.~(\ref{L1}) and Eq.~(\ref{T1}), respectively. Red (green) color corresponds to the pressure (energy) route in determining $R$; these curves are almost indistinguishable. The conventional and simplified QLCA show very good agreement in the long-wavelength regime (for $q\lesssim 2$).}
\label{Fig1}
\end{figure}

Thermodynamics of Yukawa systems has been extensively investigated and the accurate data for $u_{\rm ex}$ and $p_{\rm ex}$ from MC and MD simulations exist (see, for example, Refs.~\onlinecite{MeijerJCP1991,HamaguchiPRE1997,CaillolJSP2000}). Various fitting formulas, based on different physical arguments, have been also proposed.~\cite{RTScaling,RosenfeldPRE2000,KhrapakISM,KhrapakJCP2015,ToliasPoP2015,
Khrapak2015nearOCP,KhrapakPPCF2015} In this study we use simple practical expressions for $u_{\rm ex}$ and $p_{\rm ex}$ applicable in a wide parameter regime characterizing Yukawa fluids, which have been published recently.~\cite{KhrapakPRE2015} These expressions are essentially based on the Rosenfeld-Tarrazona freezing-temperature scaling for the thermal component of the excess internal energy and related thermodynamic quantities for simple fluids with soft repulsive interactions.~\cite{RTScaling,RosenfeldPRE2000} They demonstrate excellent agreement with the results from numerical simulations in the regime $\kappa\lesssim 5$ and $\Gamma/\Gamma_{\rm m}\gtrsim 0.1$, which is addressed in this study.

The quantities $R_u(\kappa,\Gamma)$ and $R_p(\kappa,\Gamma)$  evaluated using the expressions for $u_{\rm ex}(\kappa,\Gamma)$ and $p_{\rm ex}(\kappa,\Gamma)$ from Ref.~\onlinecite{KhrapakPRE2015} exhibit the following properties. 
For a given pair of $\kappa$ and $\Gamma$, $R_u$ is slightly larger than $R_p$. Both, $R_u$ and $R_p$ demonstrate slow increase as $\Gamma$ and $\kappa$ increase. In a relatively wide parameter regime investigated ($1\leq \kappa \leq 4$ and $0.01\leq \Gamma/\Gamma_{\rm m}\leq  1$) the values of $R$ are confined to a relatively narrow range $1\lesssim R_{u,p}\lesssim 1.3 $. An example of calculated dispersion relations using the simple model form of $g(r)$ are shown in Fig.~\ref{Fig1}. Here the solid curves correspond to the full QLCA approach with the RDF obtained from direct MD simulations (see below for description), while the dashed (longitudinal mode) and dotted (transverse mode) curves correspond to the proposed simplification employing the simplest model RDF, linked to the thermodynamic properties of the system. Two important observations are: (i) The simplified approach is practically insensitive to whether energy or pressure route  is used to determine $R$, and (ii) The simplified QLCA calculation demonstrates very good agreement with the full QLCA in the regime of sufficiently long wavelength, $q\lesssim 2$ (where the first maximum of the longitudinal wave dispersion occurs).    

\section{Detailed comparison}

Further MD simulations and full QLCA calculations have been performed to verify the main conjecture of this study. Simulations have been performed on graphics processing unit (NVIDIA GTX 960) using the HOOMD-blue software.~\cite{HOOMD,Anderson2008} We used $N=50653$ Yukawa particles in a cubic box with periodic boundary conditions. The cutoff radius for the potential has been chosen to be $L_{\rm cut}=14.5\lambda$. 
The numerical time step was set to $\simeq 10^{-2}\sqrt{ma^3/Q^2}\sim 10^{-2}\omega_{\rm p}^{-1}$. Simulations have been performed in the canonical ensemble ($NVT$) with the Langevin thermostat at a temperature corresponding to the desired target coupling parameter $\Gamma$.


\begin{figure}
\includegraphics[width=7.5cm]{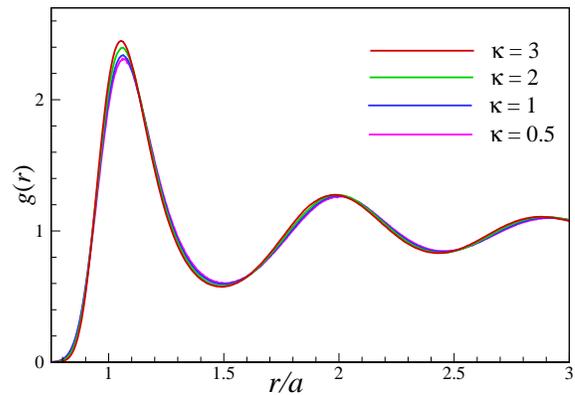}
\caption{(color online) The radial distribution functions $g(r)$ for four state points of strongly coupled Yukawa fluids, listed in Table~\ref{Tab1}. These state points are characterized by approximately the same distance from the melting curve measured in terms of the reduced coupling parameter, $\Gamma/\Gamma_{\rm m}\simeq 0.8$.  }
\label{Fig2}
\end{figure}

The system was first equilibrated for one and a half million time steps and then we saved the particle positions and trajectories every 60 time step for 80000 time steps. The particle current were then calculated
$${\bf J}({\bf k},t)=\sum_{j=1}^N {\bf v}_j(t) \exp(i {\bf k} \cdot {\bf r}_j(t))$$
and the Fourier transform in time was performed to obtain 
the current fluctuation spectra. Moreover, the particle position were saved every 4000 time step for an extra 3 million time steps to extract the accurate RDF. 

\begin{table}
\caption{\label{Tab1} The longitudinal ($c_{L}$) and transverse ($c_{T}$) sound velocities of strongly coupled Yukawa fluids evaluated using the QLCA approach (velocities are expressed in units of $\omega_{\rm p}a$). $c_{L}^{u,p}$ and $c_{T}^{u,p}$ denote the longitudinal and transverse sound velocities estimated using the simplified QLCA with the model  RDF linked to the thermodynamics via the energy and pressure route, respectively.}
\begin{ruledtabular}
\begin{tabular}{llllllll}
$\kappa$ & $\Gamma$ & $c_{L}$ &  $c_{L}^u$ & $c_{L}^p$ & $c_{T}$ & $c_{T}^u$ & $c_{T}^p$ \\ \hline
0.5 & 145 & 1.980 & 1.979 & 1.979 & 0.191 & 0.193 & 0.190  \\
1.0 & 180 & 0.959 & 0.953 & 0.956 & 0.175 & 0.174 & 0.171 \\
2.0 & 370 & 0.415 & 0.403 & 0.408 & 0.128 & 0.122 & 0.122 \\
3.0 & 990 & 0.212 & 0.202 & 0.206 & 0.081 & 0.076 & 0.076 \\
\end{tabular}
\end{ruledtabular}
\end{table}

Simulations have been performed for four pairs of $\kappa$ and $\Gamma$, which are summarized in Table~\ref{Tab1}. These points have been chosen to be located in the strongly coupled fluid state, at approximately the same distance from the melting line, $\Gamma/\Gamma_{\rm m}=T_{\rm m}/T\simeq 0.8$.  
The radial distribution functions obtained in MD simulations are shown in Fig.~\ref{Fig2}. We observe close similarity of the obtained RDFs. This observation is in line with the isomorph theory put forward recently.~\cite{GnanJCP2009,DyreJPCB2014} Isomorphs are curves in the thermodynamic phase diagram along which many properties derived from structure or dynamics are invariant in properly reduced units. The isomorph theory has been developed for liquids, which  have strongly correlated fluctuations of
their energy and pressure (referred to as Roskilde-simple or just Roskilde systems~\cite{BacherNature2014}). Yukawa fluids belong to this class and it has been recently demonstrated that the state points characterized by the same $T/T_{\rm m}$ are approximately isomorps.~\cite{VeldhorstPoP2015}  

\begin{figure}
\includegraphics[width=7.5cm]{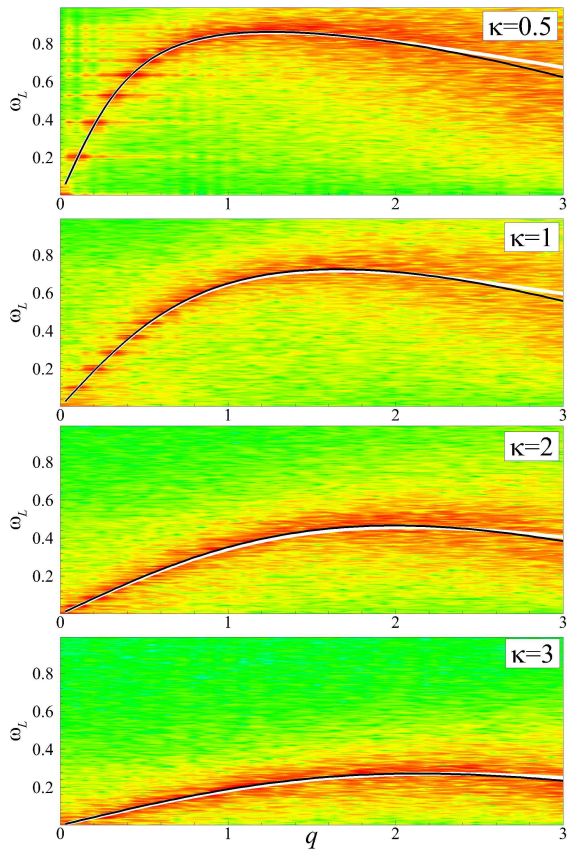}
\caption{(color online) Dispersion of the longitudinal (plasmon) mode in strongly coupled Yukawa fluids for the $(\kappa,\Gamma)$ pairs summarized in Table~\ref{Tab1}. The frequency $\omega_L$ is measured in units of the plasma frequency $\omega_{\rm p}$. The colored background corresponds to the longitudinal current fluctuation spectrum. The dark curves are the results of the full QLCA with $g(r)$ obtained using direct MD simulations. The white curves correspond to the simplified QLCA, Eq.~(\ref{L1}), with the energy route to determine $R$. The vertical scale is chosen the same in the figures to illustrate how the increase in screening (increase in $\kappa$) suppresses the wave frequency.}
\label{Fig3}
\end{figure}

\begin{figure}
\includegraphics[width=7.5cm]{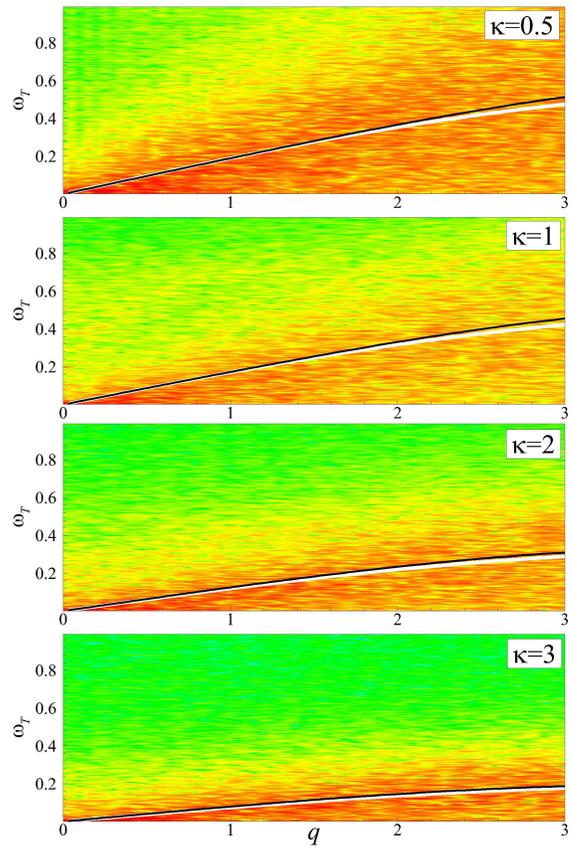}
\caption{(color online) Same as in Figure~\ref{Fig3}, but for the transverse (shear) mode.}
 \label{Fig4}
\end{figure}

Using the obtained RDFs the dispersion of the longitudinal and transverse modes within the QLCA approach have been calculated. The results are presented in Figs.~\ref{Fig3} and \ref{Fig4}. Here the color background corresponds to the spectral decomposition of the longitudinal and transverse current fluctuations. The maximum magnitude (red color) marks the approximate location of the collective excitations. The dark curves correspond to the dispersion relations calculated using the conventional (full) QLCA approach. They are in very good agreement with the current fluctuations analysis. The white curves correspond to the simplified QLCA approach discussed in the present work. In the long-wavelength regime ($q\lesssim 3$) shown in Figs.~\ref{Fig3} and~\ref{Fig4}  the agreement with the original QLCA is excellent.

In the long-wavelength limit ($q\rightarrow 0$), both longitudinal and transverse modes exhibit the acoustic dispersion, $\omega_{L/T} \simeq c_{L/T}k$. It should be reminded here that the disappearance of the shear mode at $q\rightarrow 0$ and the existence of the corresponding cutoff wave-vector $q_*$, which are well known properties of the liquid state, cannot be accounted for within the conventional QLCA, because it does not include damping effects.~\citep{DonkoJPCM2008} Nevertheless, apart from the cutoff, the QLCA shear wave dispersion appears to be nearly parallel to the actual shear wave dispersion curve and, therefore, the QLCA transverse sound velocity remains a meaningful quantity. The longitudinal ($c_L$) and transverse ($c_T$) acoustic velocities have been evaluated using the conventional QLCA as well as its simplified version using both the energy (superscript ``$u$'') and pressure (superscript ``$p$'') routes. The results are summarized in Table~\ref{Tab1}. The overall agreement is very good, the pressure route is slightly more accurate on average.
We have also estimated the thermodynamic longitudinal sound velocity using the conventional fluid approach proposed in Ref.~\onlinecite{khrapak2015fluid}. The resulting values are close but slightly lower (several percent deviation) than the QLCA approach yields, as has been already documented.~\cite{khrapak2015fluid}     

In the short-wavelength limit ($q\rightarrow \infty$), both the longitudinal and transverse frequencies approach the common limit, $\omega_{\rm E}$, the Einstein frequency which is the oscillation frequency of a single particle in the fixed environment of other particles (see Fig.~\ref{Fig1}).
For the Yukawa potential, the Einstein frequency is trivially related to the excess internal energy of the system, $\omega_{\rm E}^2/\omega_{\rm p}^2=(2\kappa^2/9\Gamma)u_{\rm ex}$. The same result can be obtained via the energy route of the present simplified QLCA [compare Eqs.~(\ref{L1}) and (\ref{T1}) in the $q\rightarrow\infty$ limit with Eq.~(\ref{u_ex})], indicating that this approach is virtually exact in the short-wavelength limit.   

\section{One-component-plasma limit} 

\begin{figure}
\includegraphics[width=7.cm]{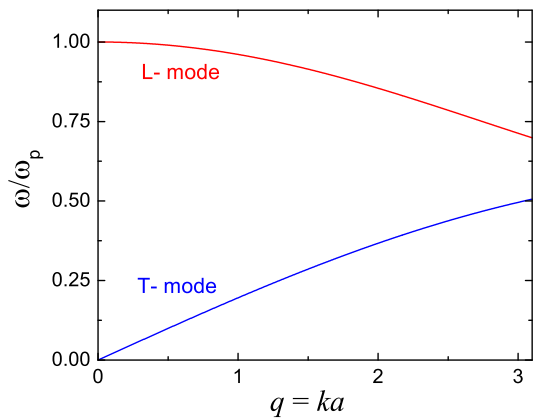}
\caption{(color online) Longitudinal (plasmon) and transverse (shear) mode dispersion curves of the one-component-plasma model. Eq.~(\ref{L2}) for the longitudinal mode and Eq.~(\ref{T2}) for the transverse mode are used, with the parameter $R=1.095$, corresponding to the strongly coupled regime. In this regime, the dispersion curves (with the normalization used) are quasi-independent on the coupling parameter $\Gamma$.}
 \label{Fig5}
\end{figure}

It is worth to briefly discuss the application of the simplified QLCA to the important limiting case of one-component-plasma (OCP). This limit corresponds to the unscreened Coulomb interaction between the particles ($\kappa=0$) and requires the presence of neutralizing background to stabilize the system and ensure finite values for the thermodynamic quantities. The dispersion relations of the simplified QLCA approach can be directly obtained from Eqs.~(\ref{L1}) and (\ref{T1}), yielding
\begin{equation}\label{L2}
\omega_L^2=\omega_{\rm p}^2\left(\frac{1}{3}-\frac{2\cos Rq}{R^2q^2}+\frac{2\sin Rq}{R^3q^3} \right)
\end{equation}
and
\begin{equation}\label{T2}
\omega_T^2=\omega_{\rm p}^2\left(\frac{1}{3}+\frac{\cos Rq}{R^2q^2}-\frac{\sin Rq}{R^3q^3} \right).
\end{equation}   
The Kohn sum rule is automatically satisfied, $\omega_L^2+2\omega_T^2=\omega_{\rm p}^2$. The energy and pressure equations have to be slightly modified due to the presence of the neutralizing background, by substituting $h(r)=g(r)-1$ instead of $g(r)$. For the Coulomb interaction the energy and pressure routes give the same result for $R$ in view of the relation $p_{\rm ex}=\tfrac{1}{3}u_{\rm ex}$. The corresponding result is $u_{\rm ex}=-\tfrac{3}{4}\Gamma R^2$. At strong coupling, the dominant contribution to the excess internal energy of the OCP model can be approximated, with a good accuracy,~\cite{DubinRMP_1999,KhrapakPoP_2014} as $u_{\rm ex}\simeq -\tfrac{9}{10}\Gamma$. Thus, in this strongly coupled regime, the parameter $R$ is practically constant, $R= \sqrt{6/5}\simeq 1.09545$. The corresponding dispersion relations are plotted in Fig.~\ref{Fig5}, the long-wavelength behavior shown here agrees well with that calculated using the conventional QLCA approach (see e.g. Fig. 4 from Ref.~\onlinecite{GoldenPoP2000}). 

In the long-wavelength limit ($q\rightarrow 0$) the longitudinal mode dispersion (\ref{L2}) reduces to
\begin{displaymath}
\frac{\omega_L^2}{\omega_{\rm p}^2} \simeq 1 -\frac{2}{30}R^2q^2= 1+\frac{4}{45}\frac{q^2u_{\rm ex}}{\Gamma}. 
\end{displaymath}
For the transverse mode we get
\begin{displaymath}
\frac{\omega_T^2}{\omega_{\rm p}^2}\simeq \frac{1}{30}R^2q^2=-\frac{2}{45}\frac{q^2u_{\rm ex}}{\Gamma}
\end{displaymath}
These expressions coincide {\it exactly} with those from the conventional QLCA approach.~\cite{GoldenPoP2000} In the opposite short-wavelength limit ($q\rightarrow \infty$) the longitudinal and transverse frequencies approach the Einstein frequency, $\omega_{\rm E}=\omega_{\rm p}/\sqrt{3}$, which again represents the exact result. 

\section{Conclusion}

To summarize, we have proposed a simplified approach to estimate wave dispersion relations in strongly coupled Yukawa fluids. The approach is based on the QLCA theory and employs the most simple model for the radial distribution function, constructed using excluded volume consideration. Analytic expressions for the longitudinal and transverse wave modes within the simplified QLCA are derived. They demonstrate very good accuracy in the long-wavelength regime and are virtually exact in the short-wavelength limit. The simplified QLCA can be useful when the exact radial distribution functions are not known, but the information about thermodynamics functions (internal energy or pressure) is available. The approach can be easily generalized to Yukawa systems in two dimensions, as well as to other interactions operating in classical systems of strongly coupled particles.   

\begin{acknowledgments}
This study was supported by the A*MIDEX grant (Nr.~ANR-11-IDEX-0001-02) funded by the French Government ``Investissements d'Avenir'' program. Structural analysis was supported by Russian Science Foundation (grant
Nr. 14-12-01185).
\end{acknowledgments}

\bibliographystyle{aipnum4-1}
\bibliography{References_KhrapakPoP2015}

\providecommand{\noopsort}[1]{}\providecommand{\singleletter}[1]{#1}%
\begin{thebibliography}{40}%
\makeatletter
\providecommand \@ifxundefined [1]{%
 \@ifx{#1\undefined}
}%
\providecommand \@ifnum [1]{%
 \ifnum #1\expandafter \@firstoftwo
 \else \expandafter \@secondoftwo
 \fi
}%
\providecommand \@ifx [1]{%
 \ifx #1\expandafter \@firstoftwo
 \else \expandafter \@secondoftwo
 \fi
}%
\providecommand \natexlab [1]{#1}%
\providecommand \enquote  [1]{``#1''}%
\providecommand \bibnamefont  [1]{#1}%
\providecommand \bibfnamefont [1]{#1}%
\providecommand \citenamefont [1]{#1}%
\providecommand \href@noop [0]{\@secondoftwo}%
\providecommand \href [0]{\begingroup \@sanitize@url \@href}%
\providecommand \@href[1]{\@@startlink{#1}\@@href}%
\providecommand \@@href[1]{\endgroup#1\@@endlink}%
\providecommand \@sanitize@url [0]{\catcode `\\12\catcode `\$12\catcode
  `\&12\catcode `\#12\catcode `\^12\catcode `\_12\catcode `\%12\relax}%
\providecommand \@@startlink[1]{}%
\providecommand \@@endlink[0]{}%
\providecommand \url  [0]{\begingroup\@sanitize@url \@url }%
\providecommand \@url [1]{\endgroup\@href {#1}{\urlprefix }}%
\providecommand \urlprefix  [0]{URL }%
\providecommand \Eprint [0]{\href }%
\providecommand \doibase [0]{http://dx.doi.org/}%
\providecommand \selectlanguage [0]{\@gobble}%
\providecommand \bibinfo  [0]{\@secondoftwo}%
\providecommand \bibfield  [0]{\@secondoftwo}%
\providecommand \translation [1]{[#1]}%
\providecommand \BibitemOpen [0]{}%
\providecommand \bibitemStop [0]{}%
\providecommand \bibitemNoStop [0]{.\EOS\space}%
\providecommand \EOS [0]{\spacefactor3000\relax}%
\providecommand \BibitemShut  [1]{\csname bibitem#1\endcsname}%
\let\auto@bib@innerbib\@empty
\bibitem [{\citenamefont {Kalman}\ and\ \citenamefont
  {Golden}(1990)}]{KaGo1990}%
  \BibitemOpen
  \bibfield  {author} {\bibinfo {author} {\bibfnamefont {G.}~\bibnamefont
  {Kalman}}\ and\ \bibinfo {author} {\bibfnamefont {K.~I.}\ \bibnamefont
  {Golden}},\ }\href {\doibase 10.1103/physreva.41.5516} {\bibfield  {journal}
  {\bibinfo  {journal} {Phys. Rev. A}\ }\textbf {\bibinfo {volume} {41}},\
  \bibinfo {pages} {5516} (\bibinfo {year} {1990})}\BibitemShut {NoStop}%
\bibitem [{\citenamefont {Golden}\ and\ \citenamefont
  {Kalman}(2000)}]{GoldenPoP2000}%
  \BibitemOpen
  \bibfield  {author} {\bibinfo {author} {\bibfnamefont {K.~I.}\ \bibnamefont
  {Golden}}\ and\ \bibinfo {author} {\bibfnamefont {G.~J.}\ \bibnamefont
  {Kalman}},\ }\href@noop {} {\bibfield  {journal} {\bibinfo  {journal} {Phys.
  Plasmas}\ }\textbf {\bibinfo {volume} {7}},\ \bibinfo {pages} {14} (\bibinfo
  {year} {2000})}\BibitemShut {NoStop}%
\bibitem [{\citenamefont {Rosenberg}\ and\ \citenamefont
  {Kalman}(1997)}]{RosenbergPRE1997}%
  \BibitemOpen
  \bibfield  {author} {\bibinfo {author} {\bibfnamefont {M.}~\bibnamefont
  {Rosenberg}}\ and\ \bibinfo {author} {\bibfnamefont {G.}~\bibnamefont
  {Kalman}},\ }\href {\doibase 10.1103/physreve.56.7166} {\bibfield  {journal}
  {\bibinfo  {journal} {Phys. Rev. E}\ }\textbf {\bibinfo {volume} {56}},\
  \bibinfo {pages} {7166} (\bibinfo {year} {1997})}\BibitemShut {NoStop}%
\bibitem [{\citenamefont {Kalman}, \citenamefont {Rosenberg},\ and\
  \citenamefont {DeWitt}(2000)}]{KalmanPRL2000}%
  \BibitemOpen
  \bibfield  {author} {\bibinfo {author} {\bibfnamefont {G.}~\bibnamefont
  {Kalman}}, \bibinfo {author} {\bibfnamefont {M.}~\bibnamefont {Rosenberg}}, \
  and\ \bibinfo {author} {\bibfnamefont {H.~E.}\ \bibnamefont {DeWitt}},\
  }\href {\doibase 10.1103/physrevlett.84.6030} {\bibfield  {journal} {\bibinfo
   {journal} {Phys. Rev. Lett.}\ }\textbf {\bibinfo {volume} {84}},\ \bibinfo
  {pages} {6030} (\bibinfo {year} {2000})}\BibitemShut {NoStop}%
\bibitem [{\citenamefont {Ohta}\ and\ \citenamefont
  {Hamaguchi}(2000)}]{OhtaPRL2000}%
  \BibitemOpen
  \bibfield  {author} {\bibinfo {author} {\bibfnamefont {H.}~\bibnamefont
  {Ohta}}\ and\ \bibinfo {author} {\bibfnamefont {S.}~\bibnamefont
  {Hamaguchi}},\ }\href {\doibase 10.1103/physrevlett.84.6026} {\bibfield
  {journal} {\bibinfo  {journal} {Phys. Rev. Lett.}\ }\textbf {\bibinfo
  {volume} {84}},\ \bibinfo {pages} {6026} (\bibinfo {year}
  {2000})}\BibitemShut {NoStop}%
\bibitem [{\citenamefont {Kalman}\ \emph {et~al.}(2004)\citenamefont {Kalman},
  \citenamefont {Hartmann}, \citenamefont {Donk{\'{o}}},\ and\ \citenamefont
  {Rosenberg}}]{KalmanPRL2004}%
  \BibitemOpen
  \bibfield  {author} {\bibinfo {author} {\bibfnamefont {G.~J.}\ \bibnamefont
  {Kalman}}, \bibinfo {author} {\bibfnamefont {P.}~\bibnamefont {Hartmann}},
  \bibinfo {author} {\bibfnamefont {Z.}~\bibnamefont {Donk{\'{o}}}}, \ and\
  \bibinfo {author} {\bibfnamefont {M.}~\bibnamefont {Rosenberg}},\ }\href
  {\doibase 10.1103/physrevlett.92.065001} {\bibfield  {journal} {\bibinfo
  {journal} {Phys. Rev. Lett.}\ }\textbf {\bibinfo {volume} {92}},\ \bibinfo
  {pages} {065001} (\bibinfo {year} {2004})}\BibitemShut {NoStop}%
\bibitem [{\citenamefont {Donko}, \citenamefont {Kalman},\ and\ \citenamefont
  {Hartmann}(2008)}]{DonkoJPCM2008}%
  \BibitemOpen
  \bibfield  {author} {\bibinfo {author} {\bibfnamefont {Z.}~\bibnamefont
  {Donko}}, \bibinfo {author} {\bibfnamefont {G.~J.}\ \bibnamefont {Kalman}}, \
  and\ \bibinfo {author} {\bibfnamefont {P.}~\bibnamefont {Hartmann}},\
  }\href@noop {} {\bibfield  {journal} {\bibinfo  {journal} {J. Phys.: Condens.
  Matter}\ }\textbf {\bibinfo {volume} {20}},\ \bibinfo {pages} {413101}
  (\bibinfo {year} {2008})}\BibitemShut {NoStop}%
\bibitem [{\citenamefont {Golden}\ \emph {et~al.}(2008)\citenamefont {Golden},
  \citenamefont {Kalman}, \citenamefont {Donko},\ and\ \citenamefont
  {Hartmann}}]{GoldenPRB2008}%
  \BibitemOpen
  \bibfield  {author} {\bibinfo {author} {\bibfnamefont {K.~I.}\ \bibnamefont
  {Golden}}, \bibinfo {author} {\bibfnamefont {G.~J.}\ \bibnamefont {Kalman}},
  \bibinfo {author} {\bibfnamefont {Z.}~\bibnamefont {Donko}}, \ and\ \bibinfo
  {author} {\bibfnamefont {P.}~\bibnamefont {Hartmann}},\ }\href {\doibase
  10.1103/physrevb.78.045304} {\bibfield  {journal} {\bibinfo  {journal} {Phys.
  Rev. B}\ }\textbf {\bibinfo {volume} {78}},\ \bibinfo {pages} {045304}
  (\bibinfo {year} {2008})}\BibitemShut {NoStop}%
\bibitem [{\citenamefont {Golden}\ \emph {et~al.}(2010)\citenamefont {Golden},
  \citenamefont {Kalman}, \citenamefont {Hartmann},\ and\ \citenamefont
  {Donk{\'{o}}}}]{GoldenPRE2010}%
  \BibitemOpen
  \bibfield  {author} {\bibinfo {author} {\bibfnamefont {K.~I.}\ \bibnamefont
  {Golden}}, \bibinfo {author} {\bibfnamefont {G.~J.}\ \bibnamefont {Kalman}},
  \bibinfo {author} {\bibfnamefont {P.}~\bibnamefont {Hartmann}}, \ and\
  \bibinfo {author} {\bibfnamefont {Z.}~\bibnamefont {Donk{\'{o}}}},\ }\href
  {\doibase 10.1103/physreve.82.036402} {\bibfield  {journal} {\bibinfo
  {journal} {Phys. Rev. E}\ }\textbf {\bibinfo {volume} {82}},\ \bibinfo
  {pages} {036402} (\bibinfo {year} {2010})}\BibitemShut {NoStop}%
\bibitem [{\citenamefont {Rosenberg}, \citenamefont {Kalman},\ and\
  \citenamefont {Grewal}(2015)}]{RosenbergCPP2015}%
  \BibitemOpen
  \bibfield  {author} {\bibinfo {author} {\bibfnamefont {M.}~\bibnamefont
  {Rosenberg}}, \bibinfo {author} {\bibfnamefont {G.~J.}\ \bibnamefont
  {Kalman}}, \ and\ \bibinfo {author} {\bibfnamefont {V.}~\bibnamefont
  {Grewal}},\ }\href {\doibase 10.1002/ctpp.201400057} {\bibfield  {journal}
  {\bibinfo  {journal} {Contrib. Plasma Phys.}\ }\textbf {\bibinfo {volume}
  {55}},\ \bibinfo {pages} {264} (\bibinfo {year} {2015})}\BibitemShut
  {NoStop}%
\bibitem [{\citenamefont {Ivlev}\ \emph {et~al.}(2012)\citenamefont {Ivlev},
  \citenamefont {Lowen}, \citenamefont {Morfill},\ and\ \citenamefont
  {Royall}}]{IvlevBook}%
  \BibitemOpen
  \bibfield  {author} {\bibinfo {author} {\bibfnamefont {A.}~\bibnamefont
  {Ivlev}}, \bibinfo {author} {\bibfnamefont {H.}~\bibnamefont {Lowen}},
  \bibinfo {author} {\bibfnamefont {G.}~\bibnamefont {Morfill}}, \ and\
  \bibinfo {author} {\bibfnamefont {C.~P.}\ \bibnamefont {Royall}},\
  }\href@noop {} {\emph {\bibinfo {title} {Complex Plasmas and Colloidal
  Dispersions: Particle-Resolved Studies of Classical Liquids and Solids
  (Series in Soft Condensed Matter)}}}\ (\bibinfo  {publisher} {World
  Scientific Publishing Company},\ \bibinfo {year} {2012})\BibitemShut
  {NoStop}%
\bibitem [{\citenamefont {Fortov}\ \emph {et~al.}(2004)\citenamefont {Fortov},
  \citenamefont {Khrapak}, \citenamefont {Khrapak}, \citenamefont {Molotkov},\
  and\ \citenamefont {Petrov}}]{FortovUFN_2004}%
  \BibitemOpen
  \bibfield  {author} {\bibinfo {author} {\bibfnamefont {V.~E.}\ \bibnamefont
  {Fortov}}, \bibinfo {author} {\bibfnamefont {A.~G.}\ \bibnamefont {Khrapak}},
  \bibinfo {author} {\bibfnamefont {S.~A.}\ \bibnamefont {Khrapak}}, \bibinfo
  {author} {\bibfnamefont {V.~I.}\ \bibnamefont {Molotkov}}, \ and\ \bibinfo
  {author} {\bibfnamefont {O.~F.}\ \bibnamefont {Petrov}},\ }\href {\doibase
  10.3367/ufnr.0174.200405b.0495} {\bibfield  {journal} {\bibinfo  {journal}
  {UFN}\ }\textbf {\bibinfo {volume} {47}},\ \bibinfo {pages} {447 } (\bibinfo
  {year} {2004})}\BibitemShut {NoStop}%
\bibitem [{\citenamefont {Fortov}\ \emph {et~al.}(2005)\citenamefont {Fortov},
  \citenamefont {Ivlev}, \citenamefont {Khrapak}, \citenamefont {Khrapak},\
  and\ \citenamefont {Morfill}}]{FortovRev}%
  \BibitemOpen
  \bibfield  {author} {\bibinfo {author} {\bibfnamefont {V.~E.}\ \bibnamefont
  {Fortov}}, \bibinfo {author} {\bibfnamefont {A.}~\bibnamefont {Ivlev}},
  \bibinfo {author} {\bibfnamefont {S.}~\bibnamefont {Khrapak}}, \bibinfo
  {author} {\bibfnamefont {A.}~\bibnamefont {Khrapak}}, \ and\ \bibinfo
  {author} {\bibfnamefont {G.}~\bibnamefont {Morfill}},\ }\href@noop {}
  {\bibfield  {journal} {\bibinfo  {journal} {Phys. Rep.}\ }\textbf {\bibinfo
  {volume} {421}},\ \bibinfo {pages} {1} (\bibinfo {year} {2005})}\BibitemShut
  {NoStop}%
\bibitem [{\citenamefont {Robbins}, \citenamefont {Kremer},\ and\ \citenamefont
  {Grest}(1988)}]{RobbinsJCP1988}%
  \BibitemOpen
  \bibfield  {author} {\bibinfo {author} {\bibfnamefont {M.~O.}\ \bibnamefont
  {Robbins}}, \bibinfo {author} {\bibfnamefont {K.}~\bibnamefont {Kremer}}, \
  and\ \bibinfo {author} {\bibfnamefont {G.~S.}\ \bibnamefont {Grest}},\ }\href
  {\doibase 10.1063/1.453924} {\bibfield  {journal} {\bibinfo  {journal} {J.
  Chem. Phys.}\ }\textbf {\bibinfo {volume} {88}},\ \bibinfo {pages} {3286}
  (\bibinfo {year} {1988})}\BibitemShut {NoStop}%
\bibitem [{\citenamefont {Hamaguchi}, \citenamefont {Farouki},\ and\
  \citenamefont {Dubin}(1997)}]{HamaguchiPRE1997}%
  \BibitemOpen
  \bibfield  {author} {\bibinfo {author} {\bibfnamefont {S.}~\bibnamefont
  {Hamaguchi}}, \bibinfo {author} {\bibfnamefont {R.~T.}\ \bibnamefont
  {Farouki}}, \ and\ \bibinfo {author} {\bibfnamefont {D.~H.~E.}\ \bibnamefont
  {Dubin}},\ }\href {\doibase 10.1103/physreve.56.4671} {\bibfield  {journal}
  {\bibinfo  {journal} {Phys. Rev. E}\ }\textbf {\bibinfo {volume} {56}},\
  \bibinfo {pages} {4671} (\bibinfo {year} {1997})}\BibitemShut {NoStop}%
\bibitem [{\citenamefont {Vaulina}\ and\ \citenamefont
  {Khrapak}(2000)}]{VaulinaJETP2000}%
  \BibitemOpen
  \bibfield  {author} {\bibinfo {author} {\bibfnamefont {O.~S.}\ \bibnamefont
  {Vaulina}}\ and\ \bibinfo {author} {\bibfnamefont {S.~A.}\ \bibnamefont
  {Khrapak}},\ }\href {\doibase 10.1134/1.559102} {\bibfield  {journal}
  {\bibinfo  {journal} {JETP}\ }\textbf {\bibinfo {volume} {90}},\ \bibinfo
  {pages} {287} (\bibinfo {year} {2000})}\BibitemShut {NoStop}%
\bibitem [{\citenamefont {Vaulina}, \citenamefont {Khrapak},\ and\
  \citenamefont {Morfill}(2002)}]{VaulinaPRE2002}%
  \BibitemOpen
  \bibfield  {author} {\bibinfo {author} {\bibfnamefont {O.}~\bibnamefont
  {Vaulina}}, \bibinfo {author} {\bibfnamefont {S.}~\bibnamefont {Khrapak}}, \
  and\ \bibinfo {author} {\bibfnamefont {G.}~\bibnamefont {Morfill}},\ }\href
  {\doibase 10.1103/physreve.66.016404} {\bibfield  {journal} {\bibinfo
  {journal} {Phys. Rev. E}\ }\textbf {\bibinfo {volume} {66}},\ \bibinfo
  {pages} {016404} (\bibinfo {year} {2002})}\BibitemShut {NoStop}%
\bibitem [{\citenamefont {Khrapak}\ and\ \citenamefont
  {Morfill}(2009)}]{KhrapakPRL2009}%
  \BibitemOpen
  \bibfield  {author} {\bibinfo {author} {\bibfnamefont {S.~A.}\ \bibnamefont
  {Khrapak}}\ and\ \bibinfo {author} {\bibfnamefont {G.~E.}\ \bibnamefont
  {Morfill}},\ }\href {\doibase 10.1103/physrevlett.103.255003} {\bibfield
  {journal} {\bibinfo  {journal} {Phys. Rev. Lett.}\ }\textbf {\bibinfo
  {volume} {103}},\ \bibinfo {pages} {255003} (\bibinfo {year}
  {2009})}\BibitemShut {NoStop}%
\bibitem [{\citenamefont {Yazdi}\ \emph {et~al.}(2014)\citenamefont {Yazdi},
  \citenamefont {Ivlev}, \citenamefont {Khrapak}, \citenamefont {Thomas},
  \citenamefont {Morfill}, \citenamefont {L\"{o}wen}, \citenamefont {Wysocki},\
  and\ \citenamefont {Sperl}}]{YazdiPRE2014}%
  \BibitemOpen
  \bibfield  {author} {\bibinfo {author} {\bibfnamefont {A.}~\bibnamefont
  {Yazdi}}, \bibinfo {author} {\bibfnamefont {A.}~\bibnamefont {Ivlev}},
  \bibinfo {author} {\bibfnamefont {S.}~\bibnamefont {Khrapak}}, \bibinfo
  {author} {\bibfnamefont {H.}~\bibnamefont {Thomas}}, \bibinfo {author}
  {\bibfnamefont {G.~E.}\ \bibnamefont {Morfill}}, \bibinfo {author}
  {\bibfnamefont {H.}~\bibnamefont {L\"{o}wen}}, \bibinfo {author}
  {\bibfnamefont {A.}~\bibnamefont {Wysocki}}, \ and\ \bibinfo {author}
  {\bibfnamefont {M.}~\bibnamefont {Sperl}},\ }\href {\doibase
  10.1103/physreve.89.063105} {\bibfield  {journal} {\bibinfo  {journal} {Phys.
  Rev. E}\ }\textbf {\bibinfo {volume} {89}},\ \bibinfo {pages} {063105}
  (\bibinfo {year} {2014})}\BibitemShut {NoStop}%
\bibitem [{\citenamefont {Rao}, \citenamefont {Shukla},\ and\ \citenamefont
  {Yu}(1990)}]{Rao1990}%
  \BibitemOpen
  \bibfield  {author} {\bibinfo {author} {\bibfnamefont {N.}~\bibnamefont
  {Rao}}, \bibinfo {author} {\bibfnamefont {P.}~\bibnamefont {Shukla}}, \ and\
  \bibinfo {author} {\bibfnamefont {M.}~\bibnamefont {Yu}},\ }\href {\doibase
  10.1016/0032-0633(90)90147-i} {\bibfield  {journal} {\bibinfo  {journal}
  {Planet. Space Sci.}\ }\textbf {\bibinfo {volume} {38}},\ \bibinfo {pages}
  {543} (\bibinfo {year} {1990})}\BibitemShut {NoStop}%
\bibitem [{\citenamefont {Hansen}\ and\ \citenamefont
  {McDonald}(2006)}]{HansenBook}%
  \BibitemOpen
  \bibfield  {author} {\bibinfo {author} {\bibfnamefont {J.~P.}\ \bibnamefont
  {Hansen}}\ and\ \bibinfo {author} {\bibfnamefont {I.~R.}\ \bibnamefont
  {McDonald}},\ }\href@noop {} {\emph {\bibinfo {title} {Theory of simple
  liquids}}}\ (\bibinfo  {publisher} {Elsevier Academic Press},\ \bibinfo
  {address} {London Burlington, MA},\ \bibinfo {year} {2006})\BibitemShut
  {NoStop}%
\bibitem [{\citenamefont {Meijer}\ and\ \citenamefont
  {Frenkel}(1991)}]{MeijerJCP1991}%
  \BibitemOpen
  \bibfield  {author} {\bibinfo {author} {\bibfnamefont {E.~J.}\ \bibnamefont
  {Meijer}}\ and\ \bibinfo {author} {\bibfnamefont {D.}~\bibnamefont
  {Frenkel}},\ }\href {\doibase 10.1063/1.459898} {\bibfield  {journal}
  {\bibinfo  {journal} {J. Chem. Phys.}\ }\textbf {\bibinfo {volume} {94}},\
  \bibinfo {pages} {2269} (\bibinfo {year} {1991})}\BibitemShut {NoStop}%
\bibitem [{\citenamefont {Caillol}\ and\ \citenamefont
  {Gilles}(2000)}]{CaillolJSP2000}%
  \BibitemOpen
  \bibfield  {author} {\bibinfo {author} {\bibfnamefont {J.~M.}\ \bibnamefont
  {Caillol}}\ and\ \bibinfo {author} {\bibfnamefont {D.}~\bibnamefont
  {Gilles}},\ }\href {\doibase 10.1023/a:1018727428374} {\bibfield  {journal}
  {\bibinfo  {journal} {J. Stat. Phys.}\ }\textbf {\bibinfo {volume} {100}},\
  \bibinfo {pages} {933} (\bibinfo {year} {2000})}\BibitemShut {NoStop}%
\bibitem [{\citenamefont {Rosenfeld}\ and\ \citenamefont
  {Tarazona}(1998)}]{RTScaling}%
  \BibitemOpen
  \bibfield  {author} {\bibinfo {author} {\bibfnamefont {Y.}~\bibnamefont
  {Rosenfeld}}\ and\ \bibinfo {author} {\bibfnamefont {P.}~\bibnamefont
  {Tarazona}},\ }\href@noop {} {\bibfield  {journal} {\bibinfo  {journal} {Mol.
  Phys.}\ }\textbf {\bibinfo {volume} {95}} (\bibinfo {year}
  {1998})}\BibitemShut {NoStop}%
\bibitem [{\citenamefont {Rosenfeld}(2000)}]{RosenfeldPRE2000}%
  \BibitemOpen
  \bibfield  {author} {\bibinfo {author} {\bibfnamefont {Y.}~\bibnamefont
  {Rosenfeld}},\ }\href {\doibase 10.1103/physreve.62.7524} {\bibfield
  {journal} {\bibinfo  {journal} {Phys. Rev. E}\ }\textbf {\bibinfo {volume}
  {62}},\ \bibinfo {pages} {7524} (\bibinfo {year} {2000})}\BibitemShut
  {NoStop}%
\bibitem [{\citenamefont {Khrapak}\ \emph {et~al.}(2014)\citenamefont
  {Khrapak}, \citenamefont {Khrapak}, \citenamefont {Ivlev},\ and\
  \citenamefont {Thomas}}]{KhrapakISM}%
  \BibitemOpen
  \bibfield  {author} {\bibinfo {author} {\bibfnamefont {S.~A.}\ \bibnamefont
  {Khrapak}}, \bibinfo {author} {\bibfnamefont {A.~G.}\ \bibnamefont
  {Khrapak}}, \bibinfo {author} {\bibfnamefont {A.~V.}\ \bibnamefont {Ivlev}},
  \ and\ \bibinfo {author} {\bibfnamefont {H.~M.}\ \bibnamefont {Thomas}},\
  }\href {\doibase 10.1063/1.4904309} {\bibfield  {journal} {\bibinfo
  {journal} {Phys. Plasmas}\ }\textbf {\bibinfo {volume} {21}},\ \bibinfo
  {pages} {123705} (\bibinfo {year} {2014})}\BibitemShut {NoStop}%
\bibitem [{\citenamefont {Khrapak}\ \emph
  {et~al.}(2015{\natexlab{a}})\citenamefont {Khrapak}, \citenamefont
  {Kryuchkov}, \citenamefont {Yurchenko},\ and\ \citenamefont
  {Thomas}}]{KhrapakJCP2015}%
  \BibitemOpen
  \bibfield  {author} {\bibinfo {author} {\bibfnamefont {S.~A.}\ \bibnamefont
  {Khrapak}}, \bibinfo {author} {\bibfnamefont {N.~P.}\ \bibnamefont
  {Kryuchkov}}, \bibinfo {author} {\bibfnamefont {S.~O.}\ \bibnamefont
  {Yurchenko}}, \ and\ \bibinfo {author} {\bibfnamefont {H.~M.}\ \bibnamefont
  {Thomas}},\ }\href {\doibase 10.1063/1.4921223} {\bibfield  {journal}
  {\bibinfo  {journal} {J. Chem. Phys.}\ }\textbf {\bibinfo {volume} {142}},\
  \bibinfo {pages} {194903} (\bibinfo {year} {2015}{\natexlab{a}})}\BibitemShut
  {NoStop}%
\bibitem [{\citenamefont {Tolias}, \citenamefont {Ratynskaia},\ and\
  \citenamefont {de~Angelis}(2015)}]{ToliasPoP2015}%
  \BibitemOpen
  \bibfield  {author} {\bibinfo {author} {\bibfnamefont {P.}~\bibnamefont
  {Tolias}}, \bibinfo {author} {\bibfnamefont {S.}~\bibnamefont {Ratynskaia}},
  \ and\ \bibinfo {author} {\bibfnamefont {U.}~\bibnamefont {de~Angelis}},\
  }\href {\doibase 10.1063/1.4928113} {\bibfield  {journal} {\bibinfo
  {journal} {Phys. Plasmas}\ }\textbf {\bibinfo {volume} {22}},\ \bibinfo
  {pages} {083703} (\bibinfo {year} {2015})}\BibitemShut {NoStop}%
\bibitem [{\citenamefont {Khrapak}\ \emph
  {et~al.}(2015{\natexlab{b}})\citenamefont {Khrapak}, \citenamefont {Semenov},
  \citenamefont {Cou{\"e}del},\ and\ \citenamefont
  {Thomas}}]{Khrapak2015nearOCP}%
  \BibitemOpen
  \bibfield  {author} {\bibinfo {author} {\bibfnamefont {S.~A.}\ \bibnamefont
  {Khrapak}}, \bibinfo {author} {\bibfnamefont {I.~L.}\ \bibnamefont
  {Semenov}}, \bibinfo {author} {\bibfnamefont {L.}~\bibnamefont
  {Cou{\"e}del}}, \ and\ \bibinfo {author} {\bibfnamefont {H.~M.}\ \bibnamefont
  {Thomas}},\ }\href@noop {} {\bibfield  {journal} {\bibinfo  {journal} {Phys.
  Plasmas}\ }\textbf {\bibinfo {volume} {22}},\ \bibinfo {pages} {083706}
  (\bibinfo {year} {2015}{\natexlab{b}})}\BibitemShut {NoStop}%
\bibitem [{\citenamefont {Khrapak}(2015)}]{KhrapakPPCF2015}%
  \BibitemOpen
  \bibfield  {author} {\bibinfo {author} {\bibfnamefont {S.~A.}\ \bibnamefont
  {Khrapak}},\ }\href@noop {} {\bibfield  {journal} {\bibinfo  {journal}
  {Plasma Phys. Control. Fusion}\ }\textbf {\bibinfo {volume} {58}},\ \bibinfo
  {pages} {014022} (\bibinfo {year} {2015})}\BibitemShut {NoStop}%
\bibitem [{\citenamefont {Khrapak}\ and\ \citenamefont
  {Thomas}(2015{\natexlab{a}})}]{KhrapakPRE2015}%
  \BibitemOpen
  \bibfield  {author} {\bibinfo {author} {\bibfnamefont {S.~A.}\ \bibnamefont
  {Khrapak}}\ and\ \bibinfo {author} {\bibfnamefont {H.~M.}\ \bibnamefont
  {Thomas}},\ }\href {\doibase 10.1103/physreve.91.023108} {\bibfield
  {journal} {\bibinfo  {journal} {Phys. Rev. E}\ }\textbf {\bibinfo {volume}
  {91}},\ \bibinfo {pages} {023108} (\bibinfo {year}
  {2015}{\natexlab{a}})}\BibitemShut {NoStop}%
\bibitem [{HOO()}]{HOOMD}%
  \BibitemOpen
  \href@noop {} {\emph {\bibinfo {title} {HOOMD-blue web page, see
  http://codeblue.umich.edu/hoomd-blue}}}\BibitemShut {NoStop}%
\bibitem [{\citenamefont {Anderson}, \citenamefont {Lorenz},\ and\
  \citenamefont {Travesset}(2008)}]{Anderson2008}%
  \BibitemOpen
  \bibfield  {author} {\bibinfo {author} {\bibfnamefont {J.~A.}\ \bibnamefont
  {Anderson}}, \bibinfo {author} {\bibfnamefont {C.~D.}\ \bibnamefont
  {Lorenz}}, \ and\ \bibinfo {author} {\bibfnamefont {A.}~\bibnamefont
  {Travesset}},\ }\href {\doibase 10.1016/j.jcp.2008.01.047} {\bibfield
  {journal} {\bibinfo  {journal} {J. Comput. Phys.}\ }\textbf {\bibinfo
  {volume} {227}},\ \bibinfo {pages} {5342} (\bibinfo {year}
  {2008})}\BibitemShut {NoStop}%
\bibitem [{\citenamefont {Gnan}\ \emph {et~al.}(2009)\citenamefont {Gnan},
  \citenamefont {Schr{\o}der}, \citenamefont {Pedersen}, \citenamefont
  {Bailey},\ and\ \citenamefont {Dyre}}]{GnanJCP2009}%
  \BibitemOpen
  \bibfield  {author} {\bibinfo {author} {\bibfnamefont {N.}~\bibnamefont
  {Gnan}}, \bibinfo {author} {\bibfnamefont {T.~B.}\ \bibnamefont
  {Schr{\o}der}}, \bibinfo {author} {\bibfnamefont {U.~R.}\ \bibnamefont
  {Pedersen}}, \bibinfo {author} {\bibfnamefont {N.~P.}\ \bibnamefont
  {Bailey}}, \ and\ \bibinfo {author} {\bibfnamefont {J.~C.}\ \bibnamefont
  {Dyre}},\ }\href {\doibase 10.1063/1.3265957} {\bibfield  {journal} {\bibinfo
   {journal} {J. Chem. Phys.}\ }\textbf {\bibinfo {volume} {131}},\ \bibinfo
  {pages} {234504} (\bibinfo {year} {2009})}\BibitemShut {NoStop}%
\bibitem [{\citenamefont {Dyre}(2014)}]{DyreJPCB2014}%
  \BibitemOpen
  \bibfield  {author} {\bibinfo {author} {\bibfnamefont {J.~C.}\ \bibnamefont
  {Dyre}},\ }\href {\doibase 10.1021/jp501852b} {\bibfield  {journal} {\bibinfo
   {journal} {J. Phys. Chem. B}\ }\textbf {\bibinfo {volume} {118}},\ \bibinfo
  {pages} {10007} (\bibinfo {year} {2014})}\BibitemShut {NoStop}%
\bibitem [{\citenamefont {Bacher}, \citenamefont {Schr{\o}der},\ and\
  \citenamefont {Dyre}(2014)}]{BacherNature2014}%
  \BibitemOpen
  \bibfield  {author} {\bibinfo {author} {\bibfnamefont {A.~K.}\ \bibnamefont
  {Bacher}}, \bibinfo {author} {\bibfnamefont {T.~B.}\ \bibnamefont
  {Schr{\o}der}}, \ and\ \bibinfo {author} {\bibfnamefont {J.~C.}\ \bibnamefont
  {Dyre}},\ }\href {\doibase 10.1038/ncomms6424} {\bibfield  {journal}
  {\bibinfo  {journal} {Nat. Commun.}\ }\textbf {\bibinfo {volume} {5}},\
  \bibinfo {pages} {5424} (\bibinfo {year} {2014})}\BibitemShut {NoStop}%
\bibitem [{\citenamefont {Veldhorst}, \citenamefont {Schr{\o}der},\ and\
  \citenamefont {Dyre}(2015)}]{VeldhorstPoP2015}%
  \BibitemOpen
  \bibfield  {author} {\bibinfo {author} {\bibfnamefont {A.~A.}\ \bibnamefont
  {Veldhorst}}, \bibinfo {author} {\bibfnamefont {T.~B.}\ \bibnamefont
  {Schr{\o}der}}, \ and\ \bibinfo {author} {\bibfnamefont {J.~C.}\ \bibnamefont
  {Dyre}},\ }\href {\doibase 10.1063/1.4926822} {\bibfield  {journal} {\bibinfo
   {journal} {Phys. Plasmas}\ }\textbf {\bibinfo {volume} {22}},\ \bibinfo
  {pages} {073705} (\bibinfo {year} {2015})}\BibitemShut {NoStop}%
\bibitem [{\citenamefont {Khrapak}\ and\ \citenamefont
  {Thomas}(2015{\natexlab{b}})}]{khrapak2015fluid}%
  \BibitemOpen
  \bibfield  {author} {\bibinfo {author} {\bibfnamefont {S.~A.}\ \bibnamefont
  {Khrapak}}\ and\ \bibinfo {author} {\bibfnamefont {H.~M.}\ \bibnamefont
  {Thomas}},\ }\href@noop {} {\bibfield  {journal} {\bibinfo  {journal} {Phys.
  Rev. E}\ }\textbf {\bibinfo {volume} {91}},\ \bibinfo {pages} {033110}
  (\bibinfo {year} {2015}{\natexlab{b}})}\BibitemShut {NoStop}%
\bibitem [{\citenamefont {Dubin}\ and\ \citenamefont
  {O'Neil}(1999)}]{DubinRMP_1999}%
  \BibitemOpen
  \bibfield  {author} {\bibinfo {author} {\bibfnamefont {D.~H.~E.}\
  \bibnamefont {Dubin}}\ and\ \bibinfo {author} {\bibfnamefont {T.~M.}\
  \bibnamefont {O'Neil}},\ }\href {\doibase 10.1103/revmodphys.71.87}
  {\bibfield  {journal} {\bibinfo  {journal} {Rev. Mod. Phys.}\ }\textbf
  {\bibinfo {volume} {71}},\ \bibinfo {pages} {87} (\bibinfo {year}
  {1999})}\BibitemShut {NoStop}%
\bibitem [{\citenamefont {Khrapak}\ and\ \citenamefont
  {Khrapak}(2014)}]{KhrapakPoP_2014}%
  \BibitemOpen
  \bibfield  {author} {\bibinfo {author} {\bibfnamefont {S.~A.}\ \bibnamefont
  {Khrapak}}\ and\ \bibinfo {author} {\bibfnamefont {A.~G.}\ \bibnamefont
  {Khrapak}},\ }\href {\doibase 10.1063/1.4897386} {\bibfield  {journal}
  {\bibinfo  {journal} {Phys. Plasmas}\ }\textbf {\bibinfo {volume} {21}},\
  \bibinfo {pages} {104505} (\bibinfo {year} {2014})}\BibitemShut {NoStop}%
\end{thebibliography}%

\end{document}